\documentstyle[prl,aps,multicol,epsf,graphicx]{revtex}

\begin{document}
\draft
\title{Studies of spin-orbit scattering in noble-metal nanoparticles
using energy level tunneling spectroscopy}
\author{J.~R.~Petta and D.~C.~Ralph}
\address{Laboratory of Atomic and Solid State Physics, Cornell
University, Ithaca, New York 14853}
\date{\today}
\maketitle

\begin{abstract}
The effects of spin-orbit scattering on discrete electronic energy
levels are studied in copper, silver, and gold nanoparticles.
Level-to-level fluctuations of the effective $g$-factor for Zeeman
splitting are characterized, and the statistics are found to be
well-described by random matrix theory predictions.
The strength of spin-orbit scattering increases with atomic
number and also varies between nanoparticles made of the same
metal.  The spin-orbit scattering rates in the nanoparticles are in
order of magnitude agreement with bulk measurements on disordered
samples.
\end{abstract}

%\par
%\vspace{2.0cm}


%\newpage

\begin{multicols} {2}
\narrowtext

Spin-orbit (SO) interactions are central to determining the
energy-level structure and dynamics of atomic nuclei,
large-atomic-number atoms, and a wide variety of solid-state systems.
They have important effects on electronic band structures \cite{AM}, they
influence the symmetry class of random-matrix theories applied to
quantum systems \cite{Beenakker}, and they are put to use in
proposals for
controllable manipulation of electron spins \cite{Datta}. A
particularly fundamental
method for probing SO interactions in solid-state systems is to
examine individual quantum energy levels in nanoparticles, both
because SO matrix elements can be measured directly and because SO
interactions produce strong modifications of the
effective $g$-factors for spin-Zeeman splitting.  Previously,
measurements of three individual energy levels in particular gold nanoparticles
showed that SO effects reduce the $g$-factor to
$g \sim 0.28-0.45$ \cite{Davidovic}, in sharp contrast with Al
particles, which ordinarily
give $g\!\approx\!2$ \cite{Ralph}.  Adding 4\% gold
to an aluminum nanoparticle also increased the SO scattering, thereby reducing
the $g$-factors in the particle to $g\sim0.5-0.8$ \cite{Salinas}. In 
this paper we report
systematic studies of SO effects in noble-metal nanoparticles covering a
range of atomic number ($Z$) -- copper, silver, and gold.
We characterize the mesoscopic fluctuations, present within single
particles, of the $g$-factors and SO matrix elements $\langle H_{\rm so}
\rangle$, and find
that
the statistics for the g-factors are well-described by recent
predictions of random matrix theory (RMT) \cite{Brouwer,Matveev}.
 From the values of the g-factors and the mean energy-level spacing
$\delta$ \cite{explain},
we extract SO scattering rates for each material, and we find
order of magnitude agreement with expectations based on
weak localization and electron spin resonance measurements.

SO interactions cause the electronic energy levels in a metal grain to
be not purely spin-up or spin-down, and as a consequence they reduce
the $g$-factor for spin-Zeeman splitting below the free-electron
value of 2.  In low-$Z$ nanoparticles for which SO scattering is
sufficiently weak (with $\langle H_{\rm so} \rangle \ll \delta$)
perturbation theory can be used to calculate the $g$-factor for
a given electronic state $n$ \cite{Sone}:
\begin{equation}
g_n=2 \Bigl(1-2\sum_{m \ne n}
{{|\langle \Psi_{m\downarrow}|\!H_{\rm so}\!|\Psi_{n\uparrow}\rangle |^2}
\over {(E_n-E_m)^2}} \Bigr) .
\end{equation}
Here the $\Psi_m$ denote the unperturbed (pure-spin) single-electron
eigenstates with
energy $E_m$.
It is expected that the SO matrix elements will vary between
different pairs of unperturbed energy levels because they depend on
the precise nature of the
wavefunctions, which are chaotic and fluctuating \cite{flucteigen}. The
local energy level spacing will also vary.  As a
result, $g_n$ should exhibit mesoscopic fluctuations for
different levels, $n$.
For heavier elements, perturbation theory ceases to be valid.
However, for large SO-interaction strengths, Brouwer
{\it et al.}\ and
Matveev {\it et al.}\ have recently employed RMT to
predict the statistical distribution $P(g)$ of $g$-factors in a
nanoparticle \cite{Brouwer,Matveev}. In \cite{Brouwer}, the strength of SO
interactions is tuned continuously, allowing numerical studies of the 
full crossover from the
Guassian orthogonal ensemble
(GOE) to the Gaussian symplectic ensemble (GSE) for the energy-level
statistics. In this crossover,
it is found that the mean value of $P(g)$ shifts to smaller g-factors
and the variance of the distribution exhibits a maximum for $\langle g
\rangle \sim$
1.2.

Our tunneling samples are fabricated in a vertical geometry through a
3--10~nm bowl-shaped  hole etched in a Si$_3$N$_4$ membrane
\cite{Ralph,Ralls} (see inset,
Fig.~1).
One electrode is made by evaporating 1750 \AA \space of Al
on top of the membrane to fill the bowl.
Following oxidation in 50 mTorr of O$_2$ for 3 minutes to form a
tunnel barrier, 5--20 \AA \space of Cu, Ag, or Au is evaporated on the
lower side of the device to form particles with radii of 3--5 nm. A
second tunnel barrier is then formed by the evaporation of 11 \AA \space
of
Al$_2$O$_3$. This is followed by 1750 \AA \space of Al to make the lower
tunneling electrode.  The
samples are screened at 4.2 K to select those whose current-voltage
($I$-$V$) curves exhibit Coulomb-staircase steps, indicating
tunneling through a single nanoparticle.  Then more detailed $I$-$V$
curves are measured in a dilution refrigerator, with all electrical
leads passing through copper-powder filters to minimize the effects
of RF radiation.

Figure 1 shows $I$-$V$ and $dI/dV$-$V$ curves for a Cu 
\linebreak
\begin{figure}
\begin{center}
\leavevmode
\epsfxsize=8.5 cm
\epsfbox{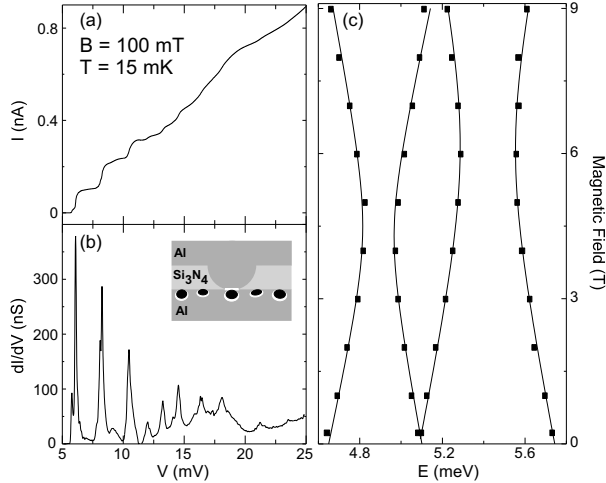}
\end{center}
\vspace{0 cm}
\caption{
\label{figure1}
(a)  $I$-$V$ and (b) $dI/dV$-$V$ curves for
Cu \#1.  (c) Magnetic field dependence of
peaks in $dI/dV$ for energy levels exhibiting
avoided level crossings.  The $V$ scale has been converted to energy, 
using the procedure discussed in the text.  Inset: Cross-sectional
device
schematic.}
\end{figure}
\noindent
device (Cu$\#$1).
As $V$ is increased, $I$ increases in discrete steps due to tunneling
through
well-resolved energy
levels within the nanoparticle \cite{Ralph}. The resonances in
the differential
conductance directly correspond to the energy level spectrum of the
nanoparticle. In an applied magnetic
field ($H$) most resonances in $dI/dV$ divide into 2 peaks, allowing
measurements of the energy-difference for spin-Zeeman splitting
between Kramers-pair eigenstates, $\Delta E = g \mu_B H$.
In order to determine the effective $g$-factors, we must convert the
magnitude of the splitting from the measured value in voltage to
energy, correcting for capacitive division of the voltage across the
two tunnel junctions according to $\Delta E = \Delta V / (1+C_1/C_2)$
where $C_1/C_2$ is the capacitance ratio for the two
tunnel junctions.  This ratio can be determined by measuring the
shift in $V$ for the resonances as the aluminum leads are changed
from superconducting to normal in small magnetic fields, or by
comparing the voltage for tunneling through the same quantum state
for positive and negative $V$ \cite{Ralph}.
The low-field $g$-factors for sample Cu$\#$1 fall in the range from
1.30$\pm$0.04 to 1.82$\pm$0.05.

In the weak SO-scattering limit where perturbation theory is valid,
the effects of SO interactions on the magnetic field evolution of
neighboring energy levels
(shown in detail in Fig.~1(c) and in broader view in Fig.~2(a)) can
be modeled by diagonalizing a simple $2 \times 2$
Hamiltonian matrix with off-diagonal matrix elements $\langle H_{\rm
so} \rangle$ coupling the
eigenstates\cite{Salinas}.  The minimum energy difference at the
avoided level crossing directly gives 2$\langle H_{\rm so}
\rangle$ between those two states.  The solid lines in
Fig.~1(c) correspond to solutions of $2 \times 2$  Hamiltonians
with $\langle H_{\rm so} \rangle$=76 $\mu$eV for the avoided crossing
near 4.9 meV, and
$\langle H_{\rm so} \rangle$=134 $\mu$eV
\linebreak
\begin{figure}
\begin{center}
\leavevmode
\epsfxsize=8.5 cm
\epsfbox{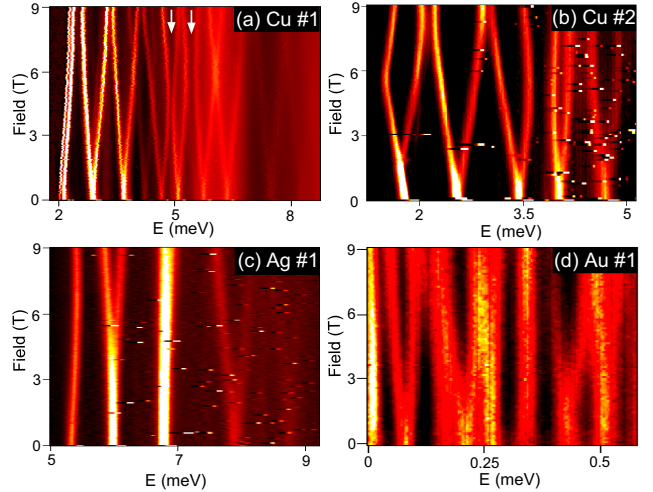}
\end{center}
%\vspace{-0.5 cm}
\caption{
\label{figure 2}
Color scale plots of $dI/dV$ in (a,b) Cu, (c) Ag,
and (d) Au samples as a function of energy and magnetic field. The 
arrows denote the avoided crossings of Fig. 1.}
\end{figure}
\noindent
for the avoided crossing near
5.4 meV.  These matrix elements are in the size range expected based
on the measured $g$-factors and the  mean level spacing $\delta =$
0.70~meV.  If the level spacing were uniform and equal to this value,
and all the matrix elements $\langle H_{\rm so} \rangle$ were $\sim$
130~$\mu$eV, then Eq.~(1) would predict $g= 2-6.6(\langle H_{\rm so}
\rangle/\delta)^2 \!\sim\!$ 1.77.
If energy-level fluctuations are included, the estimate for
%Fig. 2 was here.
the typical $g$-factor is somewhat smaller \cite{Matveev}.  In accord
with Eq.~(1), we have seen that in a given sample the states which
happen to have smaller-than-average energy spacings from neighboring
levels generally have smaller-than-average $g$-factors.

For the heavier elements Ag and Au, the g-factors become
smaller with increasing atomic number. Figure 2 shows color scale plots
of $dI/dV$ for increasing
magnetic field in Cu, Ag, and Au samples. We find $g$-factors in the
range 1.3 to 1.9 for Cu, 0.25 to
1.1 for Ag, and 0.05 to 0.19 in Au for the samples shown.
Table~1 lists the measured mean values of the g-factors and their
standard deviations for all the samples we have measured with more than three
resolvable resonances.

The most direct way to compare the statistics of the measured
$g$-factors with theoretical predictions is by plotting the
integrated probability distribution of the $g$-factors for each
sample.   In Fig.~3 the points correspond to the g-factors from
several Au, Ag, and Cu samples, while the solid lines are the theoretical
predictions from Brouwer $\textit{et al}$.\cite{Brouwer} with the
SO-scattering strength parameter $\lambda$ adjusted to
minimize the least squares error between the theoretical and
experimental distributions.
Our results are in good qualitative agreement with the RMT
predictions.  In particular, for a single value of the adjustable 
parameter $\lambda$, the
theoretical distributions account reasonably for both the mean value
of $g$ and the 
\linebreak
\begin{figure}
\begin{center}
\leavevmode
%\hspace*{-0.8cm}
%\includegraphics[height=6.5 cm, width=8.5 cm]{figure3.eps}
\epsfxsize=8.5 cm
\epsfbox{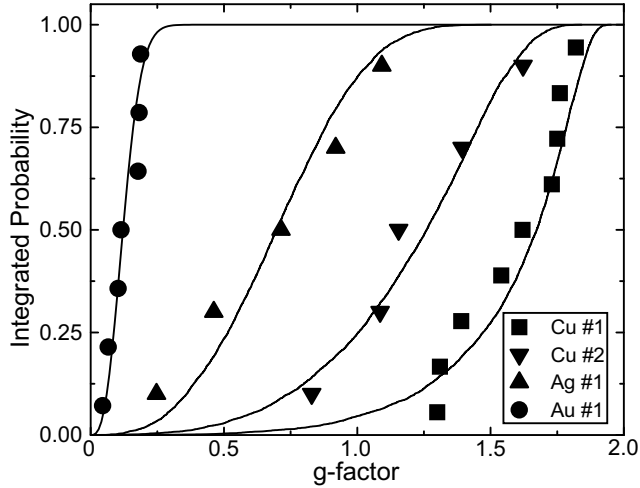}
\end{center}
%\vspace{-0.25 cm}
\caption{
\label{figure 3}
Integrated probability distributions of the $g$-factors for
several samples (points), compared to the predictions of random-matrix
theory with the SO-interaction strength adjusted for the best fit (lines). See
Table 1 for the
sample parameters.}
\end{figure}
\noindent
standard deviation for each sample.

We do observe significant differences in the average value of the
$g$-factor for different nanoparticles composed of the same material,
{\it e.g.}, for the two Cu samples depicted in Fig.~3 and for the
Ag samples \#1 and \#2.  This is true
even though the mean level spacings for the two Cu samples are nearly
identical. Because of these differences, we
cannot meaningfully consolidate the
data from many samples to improve the quality of our statistics in
making comparisons to the RMT predictions. The physical origin of SO
interactions at
low temperatures within noble metal nanoparticles is expected to be
scattering from defects or the nanoparticle surface \cite{elliot,yafet,fabian}.
Sample-to-sample variations in the strength of SO interactions are
therefore reasonable in nanoparticles, if the defect concentration or
the surface-to-volume ratio may vary, for instance due to size and
shape differences between particles.

A few of the energy levels shown in Fig.~2 do not exhibit Zeeman
splitting.  We observe this only in samples where we can identify the
tunneling transitions as corresponding to changes from an odd to an
even number of conduction electrons within the particle.  We make
this identification  by observing the lowest energy tunneling
resonance.  If the nanoparticle initially has an odd number of
electrons, then the first energy level accessible for tunneling will have
only one spin state available, and therefore the lowest energy
transition does not undergo Zeeman splitting (Fig.~2(a,c,d))
\cite{Ralph}.  In other words, the tunneling transition occurs via a
spin-singlet state. Similarly, we can explain the lack of observed
splitting in some higher energy resonances as due to tunneling via
excited spin-singlet
states.  (See the states near 7~meV in Fig.~2(c) and 0.3~meV in
Fig.~2(d).)
It is also possible that the Zeeman splitting may sometimes be sufficiently
small for a level in Ag or Au that we
cannot resolve it.
Our resolution limit is approximately $g \ge$ 0.03. From the
theoretical probability
distributions of Fig. 3, about 0.02\% of the states in Ag and 1.4\% of
the states in
Au can be expected to have g$<$0.03.

In some samples, background-charge noise and nonequilibrium effects
can interfere with accurate determinations of the statistics of
the $g$-factors.  For instance, as a function of the voltage across
the device or
as a function of time, the background charge affecting the
nanoparticle can sometimes take on an ensemble of different values,
shifting the energy levels that one is trying to measure.  When this
happens, current steps due to background-charge changes may be
misinterpreted as due to quantum states, or the same quantum state
may be measured several times at different values of voltage.  In one
Cu sample not shown in this paper, we measured six apparently
different resonances all with an identical $g$-factor of 1.50, which
we ascribe to this effect.  This behavior was not seen in any other
samples measured in this study.  If several closely-spaced levels are
observed in a sample, ways to diagnose the cause as background noise
include the presence of negative differential conductance even with
normal-state electrodes, the presence of apparent level crossings
rather than avoided crossings, and identical $g$-factors for several
resonance peaks.

An important question for understanding the properties of nm-scale
devices is whether the strength of SO scattering is consistent with
expectations based on measurements made by other techniques on larger
samples, or whether there is new physics at small scales.  Because
effective $g$-factors measured by different experimental techniques can probe
physically distinct quantities \cite{Matveev}, we will make the
comparison based on estimates for SO scattering times ($\tau_{\rm
so}$) that we can extract from our data.  We will also make
comparison only to samples with mean free paths for electron
scattering comparable to the size of our nanoparticles, since this
scattering is the probable source of SO-induced spin relaxation
\cite{elliot,fabian}.  In  the limit of strong SO scattering in a
nanoparticle, the relationship between the $g$-factor and
$\tau_{\rm so}$ has been solved analytically \cite{Matveev,explain}:
\begin{equation}
\langle g^2 \rangle= {{3}
\over {\pi \hbar}} \tau_{\rm so}\delta.
\end{equation}
For the Au samples, using the values of $\delta$ listed in Table 1, we find
$1/\tau_{\rm so}= (2 - 8) \times
10^{12} s^{-1}$.
The Ag and Cu samples fall outside the strong-scattering regime.
However by fitting to the measured $g$-factors the numerical 
distributions calculated by
Brouwer {\it et
al.}, we can still determine $\tau_{\rm so}$ for each sample from the 
single fitting parameter $\lambda$ defined in \cite{Brouwer}: 
$1/\tau_{\rm so}= \lambda^2 \delta/(\pi \hbar)$.
By this method, the SO scattering rates determined for the
two Ag samples are quite different, $1/\tau_{\rm so}= 3\times10 ^{11} s^{-1}$ and $2\times10^{12} s^{-1}$, while for Cu
$1/\tau_{\rm so}= (2 - 8)\times10 ^{11} s^{-1}$.

For Cu, we can also use perturbation theory to check the
rate determined from the RMT approach. The value of $\tau_{\rm so}$ within
perturbation theory is \cite{Halperin}
\begin{equation}
1/\tau_{\rm so}={{2\pi |\langle H_{\rm so}\rangle|^2}\over{\hbar \delta}}.
\end{equation}
For $\langle H_{\rm so} \rangle\!\sim\! 130 \space \mu$eV and $\delta$=0.7
meV, we find $1/\tau_{\rm so} \sim 2 \times 10^{11} s^{-1}$, in
agreement with the rate for Cu\#1 listed in Table 1.

These values for $\tau_{\rm so}$ can be compared to the SO
scattering rates measured
previously by weak localization, and from the linewidths of
conduction electron spin resonance (CESR) in metal foils.  From weak
localization measurements on quench-condensed noble-metal films,
Bergmann \cite{Bergmann} found $1/\tau_{\rm so} = 5.4 \times 10^{12}
s^{-1}$ for Au,
$2.6$-$3.1\times 10^{11} s^{-1}$ for Ag (depending on the disorder) and
$3.2 \times 10^{11} s^{-1}$ for Cu.  The values for Au and Cu agree
to within a factor of three with all the rates we extract, while
the weak-localization rate for Ag coincides with sample Ag\#2 but not
Ag\#1. The CESR comparisons are
less direct, because these measurements are performed on
crystalline foils with thicknesses on the order of microns or more. In
these experiments it is generally found that the spin-flip rate scales
inversely with sample thickness due to the
dominant effect of surface scattering \cite{Couch}.  The measured
values include spin-flip rates of $7 \times 10^9 s^{-1}$ for Au foils
one hundred microns thick \cite{Monod}, $1.5 \times 10^8 s^{-1}$
for a 5-$\mu$m-thick Ag foil \cite{Couch}, and $2.6 \times 10^8
s^{-1}$ for a 1.2 $\mu$m Cu film \cite{Stesmans}.  Assuming that
the rates would scale as 1/thickness down to a thickness of 5 nm, we
arrive at estimates of $1 \times 10^{14} s^{-1}$ for Au, $1 \times
10^{11} s^{-1}$ for Ag, and $6 \times 10^{10} s^{-1}$ for Cu.  These
SO scattering rates are also in rough order-of-magnitude agreement with the values
obtained from the $g$-factor analysis.

In conclusion, we have measured the $g$-factor distributions in
Cu, Ag, and Au nanoparticles using tunneling
spectroscopy.
The $g$-factors within a given sample exhibit mesoscopic
fluctuations, with distributions in good accord with
random-matrix-theory predictions.  We have observed significant
differences in both the $g$-factor distributions and $\tau_{\rm so}$
between different nanoparticles composed of
the same metal, illustrating the importance of the specific sample
structure in the origin of SO scattering.  The SO scattering rates we
extract for the nanometer-scale particles are in rough
order-of-magnitude agreement with extrapolations
from previous CESR and weak localization measurements.

We thank Piet Brouwer and Xavier Waintal for discussions
and the use of their simulation curves. This work was supported by the Packard
Foundation
and the NSF (DMR-0071631).  JRP acknowledges the support of a NSF Graduate
Research Fellowship.  Sample fabrication was performed
at the Cornell node of the National Nanofabrication Users Network.

%\linebreak
\begin{table}
\caption{
Sample parameters.  $N$ is the number of resonances resolved, 
$\langle g \rangle$ is the mean $g$-factor, $\sigma_e(g)$ is the
experimental standard deviation, $\sigma_t(g)$ is the standard 
deviation of the theory curve with the best-fit value of the SO 
parameter $\lambda$ as defined in [7], $\delta$
is the mean level spacing, and $1/\tau_{\rm so}$ is the spin-orbit 
scattering rate calculated as discussed in the text.}
\begin{center}
\begin{tabular}{|l|c|c|c|c|c|c|c|}
\ &$N$&$\langle g \rangle$&$\sigma_{e}(g)$&$\sigma_{t}(g)$& 
$\lambda$&$\delta$(meV)& 1/$\tau_{\rm so} (s^{-1})$
\\ \hline
Cu\#1&9&1.58&0.20&0.17&0.7&0.70&$2\times10^{11}$  \\
Cu\#2&5&1.22&0.30&0.31&1.2&0.71&$5\times10^{11}$  \\
Cu\#3&5&0.79&0.29&0.29&2.1&0.36&$8\times10^{11}$  \\  \hline
Ag\#1&5&0.69&0.34&0.27&2.4&0.85&$2\times10^{12}$  \\
Ag\#2&5&1.54&0.07&0.17&0.7&1.13&$3\times10^{11}$  \\  \hline
Au\#1&7&0.12&0.06&0.05&12.7&0.10&$8\times10^{12}$  \\
Au\#2&7&0.17&0.07&0.07&9.5&0.12&$5\times10^{12}$  \\
Au\#3&5&0.45&0.27&0.18&3.9&0.27&$2\times10^{12}$  \\
\end{tabular}
\end{center}
\end{table}

\end{multicols}

\begin{thebibliography}{99}
\bibitem{explain}  We define $\delta$ as the mean spacing between Kramers' doublets at $H\!=\!0$.  This is the same definition used in \cite{Brouwer}, but it is twice the level spacing defined in \cite{Matveev}.
\end{thebibliography}
\end{document}